\documentclass{article}
\usepackage{spconfa4,amsmath,graphicx}

\usepackage{multirow}
\usepackage{booktabs}
\usepackage{hyperref}
\hypersetup{hidelinks}

\newcommand{\rt}{{$\mathrm{RT}_{60}$}}

\newcommand{\dval}{{$D_{50}$}}

\title{Adapting a Text-to-Audio Model for Room Impulse Response Generation}
\name{Kirak Kim, Sungyoung Kim}
\address{Graduate School of Culture Technology, KAIST, South Korea}
\begin{document}
%
\maketitle

\begin{abstract}
Room Impulse Responses (RIRs) enable realistic acoustic simulation, with applications ranging from multimedia production to speech data augmentation. However, acquiring high-quality real-world RIRs is labor-intensive, and data scarcity remains a challenge for data-driven RIR generation approaches. In this paper, we propose a novel approach to RIR generation by adapting a pre-trained text-to-audio model, demonstrating for the first time that large-scale generative audio priors can be effectively leveraged for this task. To address the lack of text-RIR paired data, we utilize a labeling pipeline leveraging vision-language models to extract acoustic descriptions from existing image-RIR datasets. We introduce an in-context learning strategy to accommodate free-form user prompts during inference. Evaluations including a subjective listening test demonstrate that our model generates plausible RIRs with substantially less training data. Audio examples are available on our demo 
website\footnote{\href{https://kirak-kim.github.io/sao-t2r-demo/}{https://kirak-kim.github.io/sao-t2r-demo/}}.
\end{abstract}
%


\section{Introduction}
\label{sec:intro}

Room Impulse Responses (RIRs) characterize the acoustic transfer function of an enclosed space, capturing how sound propagates and interacts with the environment through reflection, absorption, and scattering. Convolving anechoic audio signal with an RIR simulates how a signal would sound within that specific space. Consequently, RIRs are utilized to create realistic auditory environments in applications ranging from multimedia production (e.g., virtual reality) to data augmentation for robust Automatic Speech Recognition (ASR). However, acquiring high-quality RIRs across diverse environments is labor-intensive, typically requiring trained experts to conduct manual measurements with specialized equipment by visiting specific environments. To address these logistical constraints, various techniques for blind room impulse response generation (generating RIRs without direct acoustic measurements of the target room) have been proposed.

Traditionally, RIR generation has relied on physics-based simulation methods. Geometric acoustic approaches, such as the Image Source Method (ISM)~\cite{10.1121/1.382599}, and ray tracing~\cite{KROKSTAD1968118}  approximate sound propagation paths, while wave-based solvers like the Finite Difference Time Domain (FDTD) method~\cite{10.1121/1.413817} numerically solve the acoustic wave equation for the given space. Scattering Delay Networks (SDNs)~\cite{7113826} synthesize reflections and reverberation using interconnected delay lines and scattering junctions. While capable of generating physically accurate RIRs, these simulation-based techniques require environmental parameters, including room geometry and surface material properties (e.g., absorption and scattering coefficients). 

To overcome the limitations of classical simulations, recent deep learning frameworks have introduced data-driven approaches that generate RIRs using various input modalities. Image-driven RIR generation models condition the generation process on room images, in which geometry of the space can be partially inferred~\cite{singh2021image2reverb, ratnarajah2024av}. While they are more convenient than classical approaches, they still require visual data of the target environment which may not always be readily accessible to general users. Alternatively, parameter-conditioned models generate RIRs directly from desired acoustic metrics~\cite{RIRAcousticParameters}. Although this eliminates the need for external data, providing plausible acoustic parameter combinations demands domain expertise. Furthermore, relying solely on acoustic parameters without geometric cues often fails to accurately capture the direct sound and early reflections of RIRs. Other approaches estimate RIRs from reference reverberant recordings~\cite{10248189, wang2025daras}, which inherently assumes the user already possesses source audio from the desired environment, limiting its practicality.

Natural language is a highly intuitive and accessible interface for conditional generation, proven by high success of Large Language Model (LLM) based services. Consequently, text-prompted RIR generation holds immense potential to implement easily usable RIR generation by eliminating the need for domain-specific inputs. PromptReverb~\cite{vosoughi2025promptreverbmultimodalroomimpulse} recently explored this by developing a text-to-RIR generation model that employs a novel captioning pipeline utilizing Vision Language Model (VLM) to address the lack of text-RIR datasets. While their approach successfully generates plausible RIRs from various natural language descriptions, they needed massive amount of RIR data (145,976 samples) to train a generative model from scratch. Although they introduced an upsampling technique based on a variational autoencoder (VAE) to accommodate RIRs with different sampling rates, the fundamental scarcity of real data still necessitated the inclusion of simulation-based synthetic RIRs. Given that synthetic RIRs have been shown to suffer from quality degradation by deviating from real-world physical constraints~\cite{10.5555/3737916.3739333}, heavily relying on them risks compromising the acoustic fidelity and accuracy of the final outputs.

Recently, large-scale generative models in the audio domain have demonstrated remarkable capabilities, capturing rich acoustic priors from diverse general audio data (e.g., speech, music, and environmental sounds)~\cite{10.1109/TASLP.2024.3399607, evans2025stable}. Recent research suggests that these learned audio priors can be effectively transferred to RIR tasks by adopting neural audio codec to RIR compression~\cite{RIRAcousticParameters, 10704165}. Inspired by this, we propose a methodology that adapts a text-to-audio (TTA) generation model on RIR data to develop a model that generates RIRs from natural language descriptions. The main contribution of this work is to demonstrate that the text understanding and audio priors of a pre-trained TTA model can be effectively leveraged to generate  plausible RIRs from only a limited amount of real-world RIR data.

\section{Proposed Method}
\subsection{Problem Definition}
Let \(c_i\) denote a natural-language description of a room and \(h_i\) its corresponding RIR. Given a paired dataset \(\mathcal{D}=\{(c_i,h_i)\}_{i=1}^{N}\), our objective is to learn the conditional distribution \(p_{\theta}(h\mid c)\), such that a plausible RIR can be generated as \(\hat{h}\sim p_{\theta}(h\mid c)\) for a previously unseen room description \(c\). In our implementation, this distribution is modeled in a latent space. The VAE encoder maps an RIR to a latent representation \(z=E(h)\), while the conditional generative model represents \(p_{\theta}(z\mid e(c))\), where \(e(c)\) denotes the text embedding. At inference, a generated latent \(\hat{z}\sim p_{\theta}(z\mid e(c))\) is converted into an RIR by the VAE decoder. This work targets blind RIR generation, which is distinct from RIR estimation tasks that infer RIRs for unseen source–receiver positions within a room~\cite{10.5555/3737916.3739333,10.5555/3600270.3600499}.

\subsection{Stable Audio Open}
Stable Audio Open (SAO)~\cite{evans2025stable} is an open source TTA generative model synthesizing non-speech audio from natural language prompts. Pre-trained on 7,300 hours of Creative Commons-licensed audio, the model is capable of generating high-fidelity 44.1kHz stereo audio from a given text prompt. The model architecture consists of three core components: 1) a pre-trained T5-base text encoder~\cite{10.5555/3455716.3455856} that extracts conditioning embeddings from input text, 2) a VAE that compresses audio waveforms into a lower-dimensional latent space and decodes them back, and 3) a Diffusion Transformer (DiT) that iteratively denoises the audio latents conditioned on the text embeddings. Functionally, the VAE encodes a general audio prior learned from large-scale pre-training and governs the acoustic fidelity of the generated waveforms, while the DiT is responsible for modeling the data distribution in the latent space conditioned on text.


\subsection{VLM-driven Dataset Labeling and Prompt Generation}
\label{VLM}
While text-RIR paired datasets are currently unavailable, several datasets provide RIRs paired with images of the corresponding room. We leverage these image-RIR datasets by employing a visual captioning approach with VLMs, similar to the recent work PromptReverb~\cite{vosoughi2025promptreverbmultimodalroomimpulse}. First, captions are generated for each room image using Qwen2.5-VL-72B-Instruct-AWQ. The VLM is prompted to act as expert acousticians, focusing explicitly on factors governing acoustic properties, such as room geometry and surface materials. Captions are refined using Gemini Flash Lite to construct the final cohesive natural language prompt, which is paired with the corresponding RIR.

Unlike PromptReverb~\cite{vosoughi2025promptreverbmultimodalroomimpulse} which generates prompts in diverse styles using VLMs in their captioning pipeline, our pipeline produces standardized prompts in a fixed format. This standardization reduces linguistic variability across text-RIR pairs, enabling more stable model training even with a relatively limited amount of paired data. To accommodate diverse, free-form natural language inputs during user inference, we incorporate an In-Context Learning (ICL) technique inspired by recent studies demonstrating the effectiveness of ICL across diverse domains~\cite{11230953}. When a user's free-form description is given, the LLM is provided with a system prompt analogous to the image captioning phase, alongside five example pairs consisting of raw captions and their corresponding refined prompts. The LLM first analyzes the input to extract relevant acoustic properties, generating an intermediate caption. Guided by the in-context examples, the LLM then translates this intermediate caption into a final standardized prompt. This approach ensures consistent and robust RIR synthesis regardless of the initial input format.

\section{Experiments}
\subsection{Experimental Setup}
We conducted our experiments using the BUT ReverbDB~\cite{szoke2019building} and OpenAIR datasets\footnote{\href{https://www.openair.hosted.york.ac.uk}{https://www.openair.hosted.york.ac.uk}}, which provide real-world RIRs paired with room images. We split the dataset in a room-disjoint manner into 1,568 training pairs from 58 rooms and 138 test pairs from 7 rooms. The training prompts were generated via the pipeline detailed in Section~\ref{VLM}.

To investigate whether priors learned from large-scale general audio data benefit RIR generation, we trained and compared two model variants: SAO-Finetuned, which initializes the DiT with pre-trained weights from the official  checkpoint\footnote{\href{https://huggingface.co/stabilityai/stable-audio-open-1.0}{https://huggingface.co/stabilityai/stable-audio-open-1.0}}, and SAO-Scratch, which initializes the DiT with random weights. In both cases, we froze the T5 text encoder and the VAE, updating only the DiT during training. This design preserves the text understanding of the text encoder and the acoustic representation prior of the VAE, while allowing the DiT to model only the probability distribution of the target RIR dataset.


The model was optimized using the AdamW optimizer (learning rate 5e-5, betas 0.9 and 0.999, weight decay 0.001) with an InverseLR scheduler. Both the SAO-Finetuned and SAO-Scratch models were trained for 50k steps (approximately 130 epochs). Training was conducted on a single NVIDIA RTX A6000 GPU and required approximately 10 hours per model.

As no publicly available text-to-RIR model exists for direct comparison, we compared our proposed method against Image2Reverb~\cite{singh2021image2reverb}, which generates RIRs given room images. This comparison is feasible as our training dataset originally comprises image-RIR pairs. We used the author's officially released weights\footnote{\href{https://media.mit.edu/~nsingh1/image2reverb/model.ckpt}{https://media.mit.edu/~nsingh1/image2reverb/model.ckpt}}  with the official inference code, as retraining the model failed with the reduced data pool used in this paper. Note that the Image2Reverb model was not originally trained on the dataset used in this study, and its training dataset partially overlaps with our test set through the inclusion of OpenAIR samples. PromptReverb~\cite{vosoughi2025promptreverbmultimodalroomimpulse} was excluded from baselines as its model weights and code are not yet publicly available.

\subsection{Quantitative Evaluation}
Following the evaluation protocols established by prior studies in the blind RIR generation task~\cite{singh2021image2reverb, vosoughi2025promptreverbmultimodalroomimpulse}, we assessed the acoustic accuracy of the synthesized RIRs by reporting the mean and median \rt\ errors in percentage. In addition, we evaluated the \dval\ (defined as the ratio of early energy within 50 ms to total energy) error to measure the difference in temporal energy structure compared to the ground-truth RIRs. As shown in Table~\ref{tab:quant_eval}, SAO-Finetuned model achieved the lowest mean \rt\ error as well as the lowest mean \dval\ errors, indicating that it most accurately reproduces both the overall reverberation time and the temporal energy distribution of the target RIRs. The SAO-Scratch variant attained the lowest \rt\ median error, suggesting that leveraging the general audio distribution trades a small amount of median \rt\ accuracy for substantially better mean-error behavior and tighter \dval\ alignment. In contrast, the Image2Reverb exhibited by far the worst performance across all metrics, likely due to its failure to generalize to room images outside its training dataset distribution. Notably, our SAO-Finetuned model achieves a mean \rt\ error of 8.4\% using only 1,568 real-world training pairs. For context, PromptReverb~\cite{vosoughi2025promptreverbmultimodalroomimpulse} reports a mean error of 8.8\% (median: -33.95\%) using 145,976 samples, approximately 93 times the size of our training set. However, these results are not directly comparable because the evaluation datasets and experimental protocols differ.

\begin{table*}[t]
\caption{Quantitative evaluation of synthesized RIRs on the test set. Values closer to 0 indicate better performance.}
\label{tab:quant_eval}
\centering
\begin{tabular}{lcccc}
\toprule
\textbf{Model} & \textbf{RT60 Mean (\%)} & \textbf{RT60 Median (\%)} & \textbf{D50 Mean (\%)} & \textbf{D50 Median (\%)} \\
\midrule
Image2Reverb & 441.3 & 204.8 & -18.4 & -22.6 \\
SAO-Finetuned & \textbf{8.4} & -53.3 & \textbf{9.4} & \textbf{7.8} \\
SAO-Scratch & 30.7 & \textbf{-42.1} & 14.1 & \textbf{7.8} \\
\bottomrule
\end{tabular}
\end{table*}

\subsection{Evaluation of In-Context Prompt Refinement}
We evaluated the effectiveness of our ICL prompt refinement pipeline using the test set. We simulated diverse user inputs by generating 50 free-form natural language prompts with distinct styles based on the ground-truth captions using Gemini 2.5 Flash, and processed them through our ICL refinement pipeline. We then extracted text embeddings via the T5 encoder of our fine-tuned SAO model and computed the cosine similarity against the ground-truth training prompt embeddings. The cosine similarity rose from $0.745$ for the raw free-form prompts to $0.926$ after ICL refinement, indicating that our ICL strategy yields T5 embeddings substantially closer to those of the training distribution. This confirms that ICL prompting effectively aligns diverse user queries with the model's training format, enabling the proposed text-to-audio model to process free-form natural language inputs in real-world RIR generation scenarios.

\subsection{Subjective Evaluation}
We conducted a subjective listening test to evaluate the perceptual quality of the generated RIRs. As comparing conditions, 1) the generated RIRs from SAO-finetuned, 2) the generated RIRs from SAO-scratch, 3) the generated RIRs from Image2Reverb model, and 4) ground-truth RIRs are convolved with clean speech utterances from the LibriSpeech~\cite{7178964} test set. During each trial, listeners were presented with the reference room image alongside the dry source signal and the reverberant stimulus. This stimulus presentation differs from prior work~\cite{vosoughi2025promptreverbmultimodalroomimpulse}, which provided VLM-generated text description of the room rather than the ground truth image itself. We adopted image-based presentation because our ground-truth dataset originally consists of (image, RIR) pairs. Evaluating against VLM-generated text would therefore conflate the perceptual quality of the generated RIRs with the accuracy of the labeling pipeline; in cases of imperfect labeling, the ground-truth RIRs and the image-conditioned baseline (Image2Reverb) would be unfairly penalized for matching the actual room rather than its potentially erroneous textual surrogate. A total of 21 listeners each evaluated 28 trials (7 rooms from the test set × 4 RIR conditions). For each trial, listeners rated the stimulus on two questions using a 5-point Likert scale (1 = not at all, 5 = very much): (Q1) spatial consistency: ``Does the speech sound as if it were recorded in the space shown in the image?'' and (Q2) audio quality: ``Does the speech sound natural and free of artifacts?'' 

\begin{figure}[t]
\centering
\includegraphics[width=0.9\columnwidth]{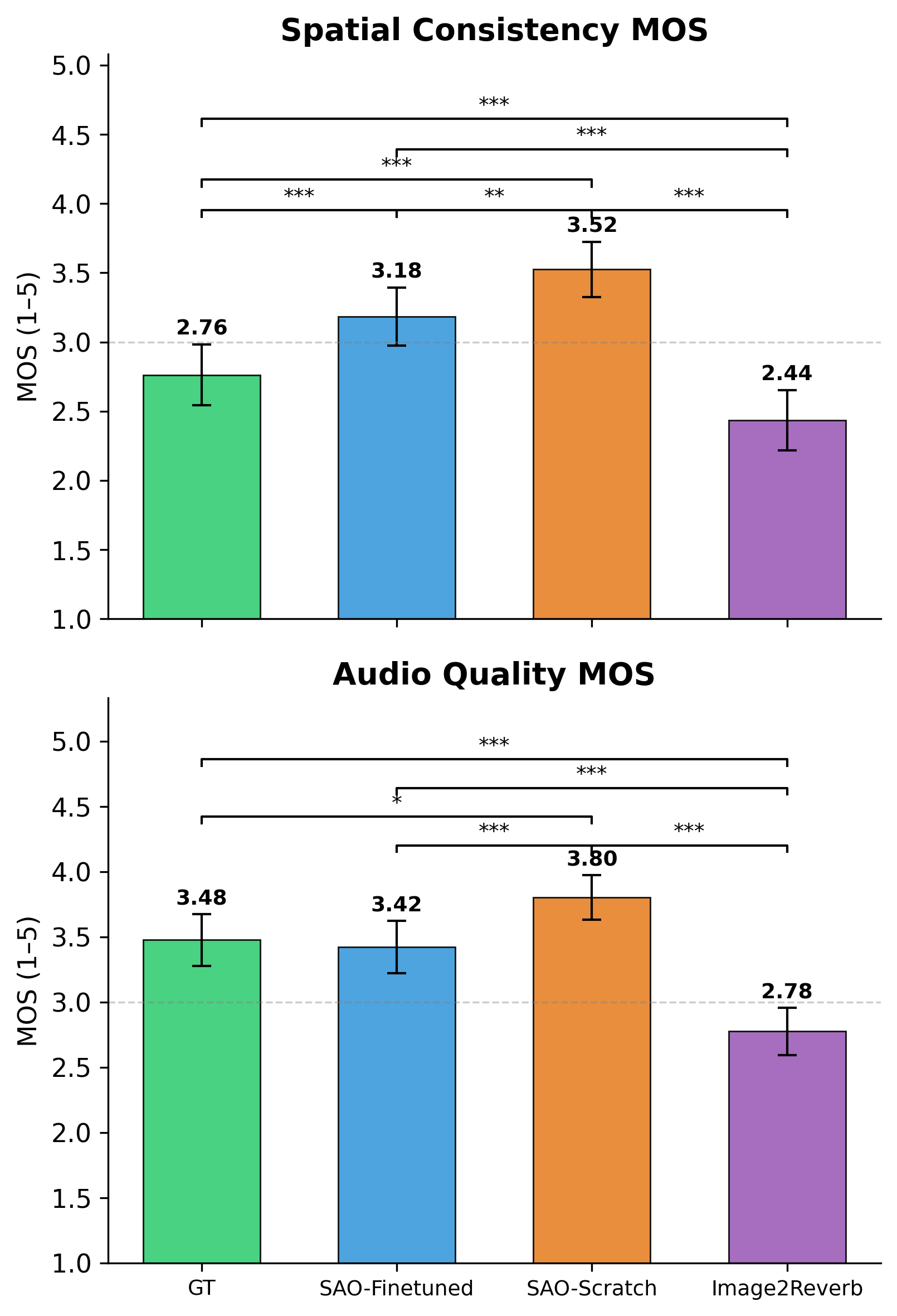}
\caption{Subjective evaluation results for spatial consistency and audio quality across different RIR generation methods.}
\label{fig:mos_overall}
\end{figure}

The listening test result is illustrated in Figure~\ref{fig:mos_overall}. A repeated-measures ANOVA with Greenhouse-Geisser correction revealed a significant main effect of method on both spatial consistency ($F(2.13, 42.52) = 52.44$, $p < .001$, $\eta_p^2 = .724$) and audio quality ($F(2.08, 41.61) = 32.15$, $p < .001$, $\eta_p^2 = .616$). Post-hoc pairwise comparisons with Bonferroni correction were 
conducted to identify which conditions differed significantly.

Both SAO-based models significantly outperformed Image2Reverb on both spatial consistency and audio quality (all $p<.001$), confirming the perceptual superiority of our text-conditioned approach over the image-conditioned baseline. More notably, both SAO-Finetuned and SAO-Scratch were rated significantly higher than the ground-truth RIRs in spatial consistency, and SAO-Scratch additionally surpassed the ground truth in audio quality (SAO-Finetuned was statistically indistinguishable from the ground truth on this measure). This is a counterintuitive outcome, given that listeners were shown the actual room images during the evaluation and the ground-truth RIRs were measured directly within those very rooms. One possible interpretation is that perceptual plausibility does not always align with physical accuracy: listeners may favor RIRs that conform to their generic expectations of how a given type of room should sound, rather than the specific acoustic signature captured at one particular source-receiver position. A rigorous test of this interpretation would require follow-up evaluation restricted to trained audio professionals, who are less likely to default to generic expectations of room acoustics (for our experiment, 7 reported formal training in audio or music, 9 identified as amateur musicians, and 5 had no related background). Image2Reverb consistently received the lowest ratings, which is consistent with its larger objective errors.

\section{Conclusion}
We present a novel text-conditioned RIR generation approach by adapting a pre-trained TTA generative model. We demonstrate for the first time that large-scale generative priors can be effectively leveraged for text-to-RIR generation. By overcoming data scarcity via leveraging audio priors and VLM-driven labeling pipeline, our model generates perceptually plausible RIRs from diverse free-form text using only a limited set of real-world RIRs. Comprehensive evaluations demonstrate its effectiveness as an accessible acoustic simulation. Despite these promising results, limitations remain. Text descriptions may not fully capture room geometry, potentially contributing to deviations from ground-truth RIRs. Explicit geometric conditioning through text-to-3D models could mitigate this limitation. Moreover, diffusion-based generation incurs iterative inference cost, which could be reduced through advanced ODE solvers or model distillation.




\section{Acknowledgments}
This work was supported by Electronics and Telecommunications Research Institute (ETRI) grant funded by the Korean government. [26CC1100, Development of User-Customized Freeform Future Display]

\bibliographystyle{IEEEbib}
\bibliography{refs}

\end{document}